%% file: main.tex
\documentclass[11pt]{article}

\usepackage{geometry}                
\usepackage{graphicx}
\usepackage{amsmath,amssymb, amsthm,epsfig,bm}
\usepackage{setspace}
\usepackage{epstopdf}
\usepackage{psfrag}
\usepackage{color,soul}
\usepackage{verbatim}
\usepackage{multirow}
\usepackage{natbib}
\usepackage{xr}
\usepackage{mathrsfs}
\usepackage{cases}
\usepackage{enumitem}

\input{style}

\newcommand{\titleshort}{A Correlation Thresholding Algorithm for Learning Factor Analysis Models}
\newcommand{\authorshort}{Kim}

\usepackage[hidelinks, unicode]{hyperref} 
\usepackage{multirow}

\numberwithin{equation}{section}

\newtheorem{theorem}{Theorem}

\theoremstyle{definition}

\usepackage{bm} 
\usepackage{graphicx} 
\usepackage{multirow} 
\usepackage[linesnumbered, vlined, ruled]{algorithm2e}
\usepackage{amssymb} 
\usepackage[colorinlistoftodos, textwidth = 20mm]{todonotes} 
\usepackage{subcaption} 
\usepackage{enumitem} 
\usepackage{amsthm}
\usepackage{appendix}
\usepackage{placeins} 
\usepackage{mathtools}
\usepackage{siunitx} 
\usepackage{booktabs}

\newcommand{\E}{\mathbb{E}}
\newcommand{\Var}{\mathrm{Var}}

\newcommand{\Prob}{\mathbb{P}}

\newcommand{\bbm}{\begin{bmatrix}}
\newcommand{\ebm}{\end{bmatrix}}

\newcommand{\thetastem}{\widehat{\theta}^{\mathrm{StEM}}_{n}}
\newcommand{\thetasir}{\widehat{\theta}^{\mathrm{SIR}}_{n_q, n}}

\begin{document}

\title{A Stochastic EM Algorithm with Sampling-Importance Resampling for Missing Data in Regression with Nonlinear Predictors}
\author{Dale S. Kim\thanks{UCLA Department of Statistics \& Data Science. Email: daleskim@stat.ucla.edu}}
\date{}
\maketitle

\begin{abstract}
Estimating regression models with nonlinear predictor transformations is challenging when data are missing, because nonlinearity typically renders the conditional distribution of the missing values intractable.
Previous methods require specific nonlinear forms, such as polynomials or interactions, or rely on approximations that can induce bias.
We propose a stochastic EM algorithm that uses sampling-importance resampling (SIR-StEM) to handle missing data under arbitrary nonlinear transformations of predictors, either substantively motivated or incidental, such as spline basis expansions.
Unlike approaches that require linearity or closed-form conditionals, SIR-StEM only requires evaluating the complete-data likelihood up to proportionality, making it applicable across a broad class of nonlinear regression models.
We construct an algorithm that makes use of missing data pattern information for computational efficiency, and establish asymptotic normality of the estimator.
We demonstrate the method in two simulation studies, one with parametric nonlinear transformations and another with a spline basis expansion based on real behavioral measures.
Results show that SIR-StEM yields low bias and near-nominal confidence interval coverage, outperforming other common approaches.
We conclude with limitations and directions for future research.

{\em Keywords:} EM algorithm, generalized linear model, nonlinear, regression, missing data.
\end{abstract}

\section{Introduction}

Nonlinear relationships are frequently hypothesized in psychological research.
Well-known examples include negative quadratic relationships between arousal and performance \citep[Yerkes-Dodson law;][]{Watters1997, Yerkes1908}, logarithmic or quadratic relationships between social interaction and well-being \citep{Ren2022}, and accelerating or decelerating effects of age on various aspects of cognition \citep{Verhaeghen1997}.

However, estimating nonlinear effects becomes challenging when data are missing.
Most missing data methods will require the conditional expectation or distribution of the missing data given the observed values, which is generally not of a well-known form, making it difficult to characterize for analysis or sampling.
Previous work has attempted to address this in several ways.
These include treating the nonlinear term as ``just another'' Gaussian random variable \citep{Seaman2012, VonHippel2009}, deriving the distribution of the missing data through numerical or analytical integration \citep{Kim2026, Ludtke2019}, or adopting Bayesian methods \citep{Kim2015, Zhang2017}.
Each of these strategies comes with limitations, such as model misspecification \citep{Bartlett2015}, poor scalability \citep{Hinrichs2014, Simonovits2003}, or requiring model-specific procedures \citep{Kim2015,Kim2026}.

\begin{figure}[!t]
\centering
\includegraphics[width=\textwidth, keepaspectratio]{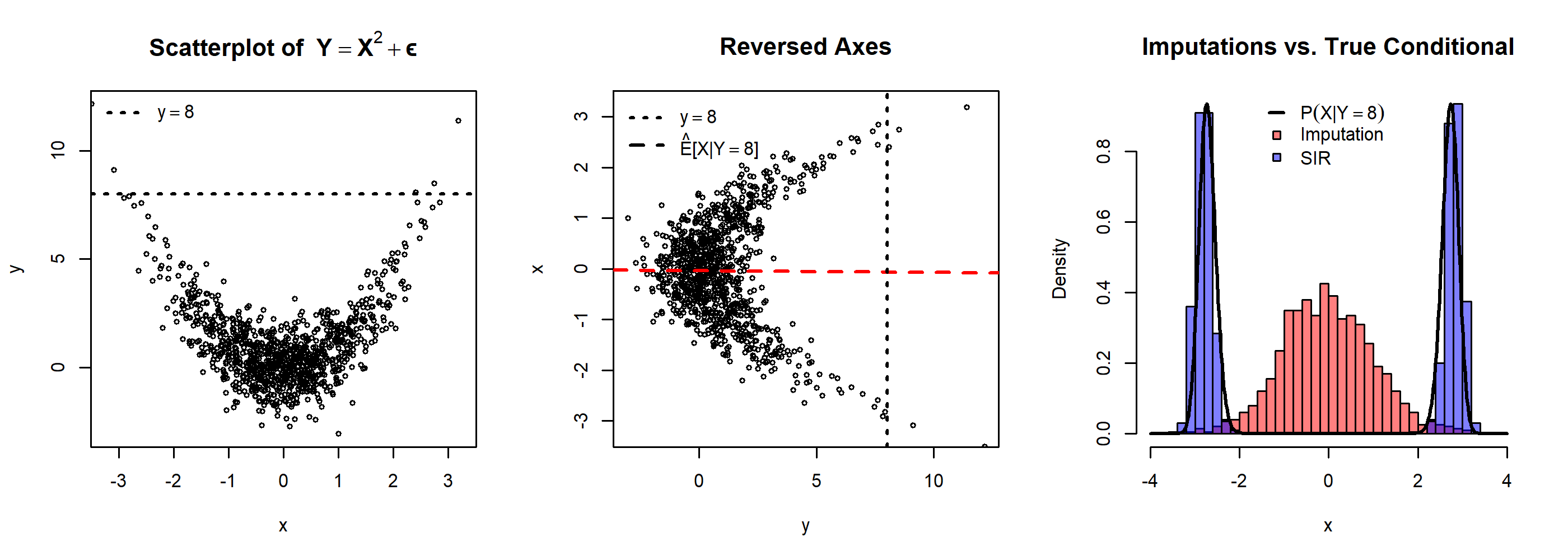} \\
\caption{A scatterplot of $Y = X^2 + \epsilon$ is displayed on the left with the dotted black line denoting $y = 8$.
In the middle, we display the scatterplot rotated, with a linear regression line $\hat{E}[X \mid y = 8]$ denoted by the dotted red line, and the dotted black line still denoting $y = 8$.
On the right, we display histograms of conditional samples $\tilde{x} \sim \Prob(X \mid y = 8)$ from the sampling-importance resampling (SIR) and linear Gaussian imputations. The solid black line denotes the true conditional distribution $\Prob(X \mid y = 8)$.}
\label{fig:example}
\end{figure}

To illustrate the core difficulty with nonlinear terms, consider a simple quadratic model: $Y = X^2 + \epsilon$, where $X$ and $\epsilon$ are standard Gaussian.
We display a scatterplot of this model in Figure~\ref{fig:example} (left).
Suppose we have a data point where $y = 8$ but the value for $x$ is missing.
Most missing data methods will require either a sample from $\Prob(X \mid y=8)$ or the expected value $\E[X \mid y=8]$.
A typical linear imputation model is shown in Figure~\ref{fig:example} (middle, red dotted line), where $Y$ is used to make a linear prediction on $X$.
The black dotted line again reflects where $y = 8$ and the intersection is where the imputations will be generated, which can be seen to be unrepresentative of the rest of the data.
This is due to the fact that the conditional distribution is bimodal, as shown in Figure~\ref{fig:example} (right, black line), meaning that methods that ignore the nonlinear structure, such as linear Gaussian imputations (shown in the red histogram), will fail to recover the true distribution.
This example highlights that even in a simple nonlinear setting, missing data problems require careful consideration, underscoring the need for estimation procedures to approximate conditional distributions that arise from nonlinear models.

To address these issues, we propose a stochastic EM algorithm using a sampling-importance resampling scheme, which can handle missing data in predictors with arbitrary nonlinear transformations.
These include substantively motivated transformations such as quadratic or interaction terms, as well as incidental transformations such as spline basis expansions.
This method is previewed in Figure~\ref{fig:example} (right, blue histogram), which can be seen to represent the bimodal conditional distribution of our previous example very well.
In the remaining sections, we establish the model and notation, provide the framework of our algorithm and its theoretical properties, and provide two empirical experiments using both simulated and real data.
We conclude with a discussion of our results and future directions.

\section{Model and Notation}

Let $X \sim \mathcal{N}_p(\mu, \Sigma)$ denote a $p$-dimensional random vector of substantive predictor variables.
Then we formulate a generalized linear regression model for a random scalar outcome variable $Y$ as follows:
\begin{equation} \label{eq:model}
\begin{aligned}
\E[Y \mid X] &= \mu_Y(X) \coloneqq g^{-1}(\phi(X)^T \beta)\\
Y \mid X &\sim \mathcal{D}_{Y \mid X}(\mu_Y(X), \psi),
\end{aligned}
\end{equation}
where $g(\cdot)$ is an invertible link function, $\beta \in \mathbb{R}^d$ is a vector of regression coefficients, and $\psi$ dnotes any additional parameters required for the conditional distribution $\mathcal{D}_{Y \mid X}(\cdot)$, such as dispersion.
We explicitly consider a nonlinear function of the substantive predictor variables in $\phi(\cdot): \mathbb{R}^p \rightarrow \mathbb{R}^d$ which maps the substantive variables ($X$) into a vector of nonlinear regressors.
For example, we may posit a relationship such as
\begin{equation}
\phi(X) = \left(1,\,\exp(X_1),\,X_2^2,\,X_3,\,X_4,\,X_3 X_4 \right),
\end{equation}
where $X$ may be variables of substantive behavioral interest, and $\phi(X)$ represents the mechanism by which $X$ relates to $Y$, or a set of nonlinear fitting functions such as splines.
Without loss of generality, we may assume that the vector $X$ contains the substantive variables for the regression model as well as any desired auxiliary variables that are useful for predicting missingness.

\subsection{Missing Data Assumptions}

At times, it will be more useful to view the data in a missing-observed distinction, rather than a predictor-outcome distinction.
To do so, we will denote a data vector as $U = (Y,\,X)$, which can be reordered as $(U_O,\,U_M)$, where $O$ is the index set of observed variables and $M$ is the index set of missing variables.
We will use lowercase $o$ and $m$ to denote the situations where the missingness pattern is realized.
The probability density/mass function of $U$ is denoted $f_{\theta}(U)$, where $\theta \in \Theta$ denotes the set of parameters for the data.
Missingness is represented by a binary indicator $R \in \{0, 1\}^{p + 1}$, which indicates whether the elements of $U$ are observed, and has a probability distribution parameterized by $\omega \in \Omega$.

We assume the data are missing at random \citep[MAR;][]{Rubin1976}, which implies that $R$ only depends on the observed data.
Formally, this is written as:
\begin{equation}
\Prob_{\omega}(R \mid U) = \Prob_{\omega}(R \mid U_O).
\end{equation}
We also assume that the values of $\theta$ and $\omega$ are distinct, or that the joint space of $\Theta$ and $\Omega$ is simply their Cartesian product $\Theta \times \Omega$.
The missing data mechanism is called ignorable if MAR and the distinctness of $\Theta$ and $\Omega$ hold \citep{Schafer1997}.

\section{The Stochastic EM Algorithm} \label{sec:em_alg}

To estimate the nonlinear model in Equation~\ref{eq:model} in the presence of missing data, we will use a stochastic expectation-maximization (StEM) algorithm \citep{Celeux1985, Diebolt1995}.
The StEM algorithm is a generalization of the traditional EM algorithm, where the \textit{E}-step is replaced with a stochastic draw.
This avoids the analytical integration necessitated by the \textit{E}-step, which can be intractable for our setting.

To begin, we briefly review the basics of the traditional EM algorithm.
The EM algorithm is a two-step iterative procedure for obtaining parameter estimates for models with missing data \citep{Dempster1977}.
Let $u_1, \dots, u_n$ be $n$ iid samples of data.
The steps are as follows:

\noindent \textbf{\textit{E}-Step.} For any iteration $t$, and corresponding parameter value $\theta^{(t)}$, define the $Q$-function as:
\begin{equation}
\begin{aligned}
Q_{\theta^{(t)}}(\theta) &= \sum_{i = 1}^n \E_{\theta^{(t)}} \left[ \log f_{\theta}(U_i) \mid u_{i,o} \right] \\
&= \sum_{i = 1}^n \int_{u_{i,m}} \log f_{\theta}(U) f_{\theta^{(t)}}(u_{i,m} \mid u_{i,o}) \, du_{i,m}.
\end{aligned}
\end{equation}

\noindent \textbf{\textit{M}-Step. } Maximize the $Q$-function with respect to $\theta$ and set the result as $\theta^{(t+1)}$:
\begin{equation}
\theta^{(t + 1)} = \underset{\theta}{\text{argmax}} \, Q_{\theta^{(t)}}(\theta).
\end{equation}
This is an iterative procedure that maximizes the expectation of the complete-data log-likelihood, given the observed data.
It is known to converge to a local maximum of the likelihood function under very general conditions \citep{Wu1983}.
Further, standard errors can be obtained by numerically differentiating the EM iterations \citep{Meng1991} or the Fisher score function \citep{Jamshidian2000}, or by parametric bootstrap \citep{Diebolt1995}.

In our nonlinear setting, the \textit{E}-step may be intractable to derive.
To address this, a stochastic version of the EM algorithm may be used \citep{Celeux1985, Diebolt1995} that instead requires a sample of the missing data given the observed data.
The idea is to replace the \textit{E}-step with a sampling step from $\mathbb{P}_{\theta^{(t)}}(U_M \mid U_O)$, then conduct the \textit{M}-Step as if the data were complete.
Iterating in this manner results in a sequence of $\theta^{(t)}$ that forms a Markov chain, with a stationary distribution that is asymptotically normal, with a mean at the maximum likelihood estimate, under mild regularity conditions \citep{Nielsen2000}.
This greatly simplifies the \textit{M}-step as closed-form maximizers and fast existing software packages for complete data can be used.
This leads to the following steps:

\noindent \textbf{Stochastic \textit{E}-Step.} At iteration $t$, for each observation $i$, sample the missing data given the observed:
\begin{equation}
\widetilde{u}^{(t)}_{i,m} \sim f_{\theta^{(t)}}(U_{i,m} \mid u_{i,o})
\end{equation}
and form the completed data vector $\widetilde{u}^{(t)}_i = (u_{i,o},\, \widetilde{u}^{(t)}_{i,m})$.

\noindent \textbf{\textit{M}-Step. } Maximize the complete-data log-likelihood as
\begin{equation}
\theta^{(t + 1)} = \underset{\theta}{\text{argmax}} \, \sum_{i = 1}^n \log f_{\theta}(\widetilde{u}^{(t)}_i).
\end{equation}
These steps are iterated until convergence, which can be assessed by the usual suite of MCMC diagnostic criteria \citep[see][for reviews]{Cowles1996, Gelman2014, Liu2008}.
In the setting of our problem, the main challenge lies in obtaining the sample $\widetilde{u}^{(t)}_{i,m} \sim f_{\theta^{(t)}}(U_{i,m} \mid u_{i,o})$, which will use a different conditional distribution depending on the missing data pattern.
We describe our strategy in the next section.

\section{The SIR-StEM Algorithm}

The remaining challenge in applying the StEM algorithm is generating samples from the conditional distribution of the missing data.
The form of this distribution depends on whether $Y$ is observed or missing, each of which requires a different sampling strategy.
As we will show, when $Y$ is missing, simple direct sampling techniques can be applied.
However when $Y$ is observed, an intractable normalizing constant is introduced.
To handle the latter situation, we adopt the sampling-importance resampling technique.

\subsection{Direct Sampling} \label{sec:samp_strat}

Direct sampling is applicable when $Y$ is missing.
First, consider the case when $Y$ is missing and all predictors are observed.
Generating samples is straightforward using:
\begin{equation}
\widetilde{Y}_i \sim \mathcal{D}_{Y \mid X}(\mu_Y(X_i), \psi).
\end{equation}
Second, if $Y$ is missing as well as some variables in $X$, then the distribution $\Prob(Y_i, X_{i,m}  \mid  X_{i,o})$ can be characterized in a generative way as $\Prob(Y_i  \mid  X_{i,m}, X_{i,o})\Prob(X_{i,m}  \mid  X_{i,o})$.
This leads to the straightforward sampling procedure of:
\begin{equation}
\begin{aligned}
\widetilde{X}_{i,m} &\sim \mathcal{N}(\mu_{i,c}, \Sigma_{i,c})\\
\widetilde{Y}_i &\sim \mathcal{D}_{Y \mid X}(\mu_Y(\widetilde{X}_i), \psi),
\end{aligned}
\end{equation}
where $\widetilde{X}_i = (X_{i,o}, \widetilde{X}_{i,m})$, and
\begin{equation}
\begin{aligned}
\mu_{i,c} &= \mu_m + \Sigma_{mo} \Sigma_o^{-1}(x_{i,o} - \mu_o) \\
\Sigma_{i,c} &= \Sigma_m - \Sigma_{mo} \Sigma_o^{-1} \Sigma_{om},
\end{aligned}
\end{equation}
which follow from the well-known conditional Gaussian properties.

\subsection{SIR Sampling}

When the outcome $Y$ is observed, direct sampling is generally no longer possible.
To see this, notice that we require samples from $\Prob(X_{i,m}  \mid  y_i, x_{i,o})$.
An analytic derivation would require the expression for
\begin{equation}
f(X_{i,m}  \mid  y_i, x_{i,o}) = \dfrac{f(y_i, x_{i,o}, X_{i,m})}{\int_{x_{i,m}} f(y_i, x_{i,o}, x_{i,m})\,d x_{i,m}},
\end{equation}
where the nonlinear relationship between $Y$ and $X_M$ will typically make the denominator intractable.
To circumvent this, we propose using the sampling-importance resampling \citep[SIR;][]{Rubin1987, Smith1992} method, which only requires knowledge of the target distribution up to a normalization constant.

The SIR algorithm works as follows.
Let $\pi(U) = ch(U)$ be the target density from which samples are desired, however only $h(U)$ is known (i.e., the normalizing constant $c$ is unknown).
Suppose we are able to sample from and evaluate a proposal density $q(U)$, which has the same support as $\pi(U)$.
Then we may emulate samples from $\pi(U)$ with the following steps:
\begin{enumerate}
  \item Draw $n_q$ proposal samples $\{u_1,\dots, u_{n_q}\}$ from $q(U)$.
  \item Calculate importance weights $w(u_i) = h(u_i)/q(u_i)$.
  \item Resample from $\{u_1,\dots, u_{n_q}\}$, assigning each $u_i$ a probability of $w(u_i)/\sum_{i = 1}^{n_q}w(u_i)$.
\end{enumerate}
This is essentially a parametric bootstrap followed by a weighted nonparametric bootstrap.
The samples generated from the SIR algorithm converge in distribution to the target asymptotically with $n_q$ \citep{Smith1992}.
The resampling step may be done with or without replacement, and the number of resamples can be greater than one, depending on the situation \citep{Gelman2014}.
For our application, we will be resampling a single time for simplicity.

In theory, any distribution can be used as the proposal distribution $q(U)$, so long as it has the same support as the target distribution.
However, the resampling step becomes more accurate the more similar the proposal distribution is to the target distribution \citep{Liu2008}, and may require fewer proposals.
Additionally, if the proposal distribution has heavier tails than the target, the probability of extremely large weights can be reduced, leading to more stable behavior \citep{Geweke1989}.
In our application, we simply used $f(X_{i,m}  \mid  x_{i,o})$ as a convenience proposal, with the conditional covariance inflated by a factor of 4 to obtain heavier tails.
We found that using a proposal sample size of $n_q = 100$ was both sufficiently fast and accurate with this proposal.

The SIR algorithm only requires the target density to be known up to a normalizing constant, and only that the density be evaluated after samples are drawn.
Therefore a convenient decomposition is
\begin{equation}
\begin{aligned}
f(x_{i,m}  \mid  y_i, x_{i,o}) &= \dfrac{f(y_i, x_{i,o}, x_{i,m})}{f(y_i, x_{i,o})}\\
&\propto f(y_i  \mid  x_{i,o}, x_{i,m}) f(x_{i,o}, x_{i,m}),
\end{aligned}
\end{equation}
which only requires evaluations of the conditional density of $Y \mid X$ as specified by the model and the Gaussian density.
Notably, this sampling procedure is agnostic to the specification of $\phi(\cdot)$, allowing for applicability to a wide range of regression problems.
This is in contrast with methods that require model-specific procedures \citep[e.g.,][]{Kim2015,Kim2026, Ludtke2019}.

\subsection{Implementation}

We incorporate these sampling techniques into the stochastic \textit{E}-step portion of the StEM algorithm, to construct a SIR-StEM algorithm.
The SIR-StEM algorithm operates by applying the direct sampling strategy when $Y$ is missing, and the SIR sampling strategy when $Y$ is observed.
We describe the complete algorithm in Algorithm~\ref{algo}.
Recall that in a StEM algorithm, the iterates $\{\theta^{(t)}\}$ form a Markov chain with a stationary distribution asymptotically centered at the maximum likelihood estimate \citep{Nielsen2000}.
As such, for our implementation, we adopt an empirical average estimator after discarding some number of burn-in iterates.
Given a maximum number of iterates $t_{\mathrm{max}}$ and a number of burn-in iterates $t_{\mathrm{burn}}$, we use the estimator
\begin{equation}
\hat{\theta} \coloneqq \dfrac{1}{t_{\mathrm{max}} - t_{\mathrm{burn}}} \sum\limits_{t = t_{\mathrm{burn}} + 1}^{t_{\mathrm{max}}} \theta^{(t)}.
\end{equation}
Of course, more sophisticated methods for MCMC convergence monitoring may also be used \citep{Cowles1996, Gelman2014, Liu2008}.

\begin{algorithm}
\caption{The SIR-StEM Algorithm}\label{algo}
\SetKwInOut{Input}{Inputs}
\SetKwInOut{Output}{Output}
\SetKwFor{For}{for}{}{endfor}
\SetKwRepeat{Repeat}{repeat}{}
\SetKwBlock{Begin}{repeat}{end}
\Input{Observed data $u_o$, model specification $\phi(\cdot)$, and start values $\theta^{(0)}$}
Set $t \leftarrow 0$\;
\While{$t \leq t_{\mathrm{max}}$}{
\textbf{Stochastic E-Step.} \For{$i = 1, \dots, n$\emph{:}}{
\uIf{\emph{only $y_i$ is missing}}{
Sample $\widetilde{Y}_i \sim \mathcal{D}_{Y \mid X}(\mu_Y^{(t)}(x_i), \psi^{(t)})$;
}
\uIf{\emph{$y_i$ and some $x_{i,m}$ are missing}}{
Sample $\widetilde{X}_{i,m} \sim \mathcal{N}(\mu^{(t)}_{i,c}, \Sigma^{(t)}_{i,c})$;\\
Form $\widetilde{x}_i = (x_{i,o}, \widetilde{x}_{i,m})$;\\
Sample $\widetilde{Y}_i \sim \mathcal{D}_{Y \mid X}(\mu_Y^{(t)}(\widetilde{x}_i), \psi^{(t)})$;
}
\uIf{\emph{$y_i$ is observed and some $x_{i,m}$ are missing}}{
\For{$j = 1, \dots, n_q$\emph{:}}{
Sample proposals $X^{*(j)}_{i,m} \sim q(X_{i,m})$;\\
Calculate importance weight $w(x^{*(j)}_{i,m}) = f_{\theta^{(t)}}(y_i, x_{i,o}, x^{*(j)}_{i,m}) / q(x^{*(j)}_{i,m})$;\\
}
Sample a single $\widetilde{X}_{i,m}$ from $\{x^{*(1)}_{i,m}, 
\ldots x^{*(n_q)}_{i,m}\}$, assigning each $x^{*(j)}_{i,m}$ a probability of $w(x^{*(j)}_{i,m})/\sum_{j = 1}^{n_q}w(x^{*(j)}_{i,m})$;
}
}
\textbf{M-Step.} Set $\theta^{(t + 1)} \leftarrow \underset{\theta}{\text{argmax}} \, \sum_{i = 1}^n \log f_{\theta}(\widetilde{u}^{(t)}_i)$\;
Set $t \leftarrow t + 1$;
}
\Output{Parameter Estimates $\hat{\theta} \leftarrow \dfrac{1}{t_{\mathrm{max}} - t_{\mathrm{burn}}} \sum\limits_{t = t_{\mathrm{burn}} + 1}^{t_{\mathrm{max}}} \theta^{(t)}$.}
\end{algorithm}

\subsection{Theoretical Properties} \label{sec:theoretical_properties}

Under standard regularity conditions, the SIR-StEM algorithm results in asymptotically normal estimates.
The main idea comes from the SIR samples converging in distribution to their target random variables, and sampling from the target random variables leads to known asymptotically normal results for the StEM algorithm.
We describe this formally as follows.

\begin{theorem}[Asymptotic Normality of SIR-StEM] \label{thm:asym_normal}
Let $\thetasir$ denote the SIR-StEM parameter estimate as described in Algorithm~\ref{algo} from a sample of size $n$ using $n_q$ proposal draws.
Under the regularity conditions of \citet{Nielsen2000}, and assuming that the SIR approximation $\Prob_{\theta, n_q}(U)$ converges continuously in $\theta$ to $\Prob_{\theta}(U)$ as $n_q \rightarrow \infty$, there exists a nondecreasing function $N: \mathbb{N} \rightarrow \mathbb{N}$ such that for every sequence $n_q(n)$ satisfying $n_q(n) \geq N(n)$,
\begin{equation}
\sqrt{n}(\widehat{\theta}^{\mathrm{SIR}}_{n_q(n), n} - \theta_0) \xrightarrow{d} \mathcal{N}(0, \mathcal{I}(\theta_0)^{-1} + \Gamma_{m})
\end{equation}
as $n \rightarrow \infty$, where $\mathcal{I}(\theta_0)$ is the observed-data information matrix and $\Gamma_{m}$ is additional asymptotic variance due to averaging a finite $m$ Markov chain, where $m = t_{\mathrm{max}} - t_{\mathrm{burn}}$ is the number of post-burn-in iterates.
\end{theorem}
The proof can be found in Appendix~\ref{app:stemsir}.
Essentially, so long as $n_q$ increases sufficiently fast with respect to $n$, SIR-StEM inherits the asymptotic distribution of the StEM algorithm.
For inference, $\mathcal{I}(\theta_0)$ may be estimated by a parametric bootstrap of the complete data using Louis' Identity \citep{Louis1982, Diebolt1995}, and $\Gamma_{m}$ can be estimated using the post-burn-in iterates using the batch means method of \citet{Jones2006}.

\section{Empirical Studies}

We now empirically evaluate the performance of our SIR-StEM algorithm in two data analysis scenarios.
We study: (1) a simulation study using parametric nonlinear transformations of predictors, varying several characteristics of the data and (2) a simulation study using real data with a spline transformation on real behavioral measures.

\subsection{Parametric Transformation of Predictors}

For our first simulation study, we sought to study estimator performance over several settings:
\begin{itemize}
  \item Estimation methods: SIR-StEM, multiple imputation (MI), predictive mean matching (PMM), and ``just another variable'' (JAV).
  \item Sample size ($n$): 250, 500, and 1000.
  \item Proportion of missingness ($\varphi_{\mathrm{MIS}}$): 0.10, 0.20, and 0.30.
  \item $R^2$ of the model: 0.25, 0.50.
  \item $r = 100$ replications per condition combination.
\end{itemize}
For the SIR-StEM algorithm we used $n_q = 100$ SIR samples and ran the Markov chain for $t_{\mathrm{max}}= 1000$ iterations.
We took the average of the latter half of the chain as our estimate.
For the proposal distribution $q(\cdot)$, we used the conditional Gaussian $f(X_{i,m} \mid x_{i,o})$ with its covariance inflated by a factor of 4 to obtain heavier tails.
For the standard errors, we used the parametric bootstrap (1000 draws) and batch mean methods described following Theorem~\ref{thm:asym_normal} for our SIR-StEM algorithm.
The multiple imputation methods used 20 imputations and their variance estimates were calculated by Rubin's rules \citep{Rubin1987b}.

\subsubsection{Data Generation} \label{sec:data_generate}

The data were generated in the following way:
\begin{equation}
\begin{aligned}
X &\sim \mathcal{N}_2(0, \Sigma)\\
\phi(X) &= (1, \exp(X_1), X_2^2)\\
Y &= \phi(X)^T \beta + \epsilon
\end{aligned}
\end{equation}
where $\Sigma$ had a diagonal of ones and off-diagonals of 0.5, each entry of $\beta$ was set to one, and $\epsilon \sim \mathcal{N}(0, \sigma^2)$, where $\sigma^2$ was chosen such that the $R^2$ of the model held per condition.

Once $X$ and $Y$ were generated, the observed data indicator $R$ was generated under a MAR mechanism.
We let $Y$ be an always observed variable and had the missingness of $X$ depend on $Y$ since that is our stochastic EM algorithm's most challenging missing data pattern.
Using a latent propensity variable $L$, we determined the cases in $X$ that were missing as a function of $Y$ according to the $\varphi_{\mathrm{MIS}}$ parameter.
This was done by setting
\begin{equation} \label{eq:mar}
\begin{aligned}
L &= \alpha_0 + \alpha_1 \left(\dfrac{y - \overline{y}}{\text{sd}(y)}\right) + \nu\\
R &= \begin{cases} 1, &\text{if } L < 0\\
0, &\text{otherwise,}
\end{cases}
\end{aligned}
\end{equation}
where $\nu\sim\mathcal{N}(0, 1 - \alpha_1^2)$.
This corresponds to a probit model for $R \mid Y$ with $\mathrm{Cor}(L, Y) = \alpha_1$, which we set to 0.25.
The desired overall missing data rate can be approximated by setting $\alpha_0$ to $\Phi^{-1}(\varphi_{\mathrm{MIS}})$, where $\Phi(\cdot)$ is the standard Gaussian CDF.
Both variables in $X$ had identically distributed latent propensities of missingness.

\subsubsection{Performance Metrics}

To evaluate performance, we calculated mean relative bias and coverage rates of the 95\% confidence interval as follows:
\begin{equation} \label{eq:outcomes}
\begin{aligned}
\text{Mean Relative Bias} &\coloneqq \dfrac{1}{r} \sum_{i = 1}^r \dfrac{(\hat{\beta}_{ij} - \beta_{j})}{\beta_j}\\
\text{Coverage Rate} &\coloneqq \dfrac{1}{r} \sum_{i = 1}^r I\left(\lvert\hat{\beta}_{ij} - \beta_j\rvert \leq \Phi^{-1}(0.975) \sqrt{\widehat{\Var}(\hat{\beta}_{ij})}\right),
\end{aligned}
\end{equation}
where $I(\cdot)$ is a binary indicator function.

\subsubsection{Results}

\begin{figure}[!th]
\centering
\includegraphics*[width=\textwidth, keepaspectratio]{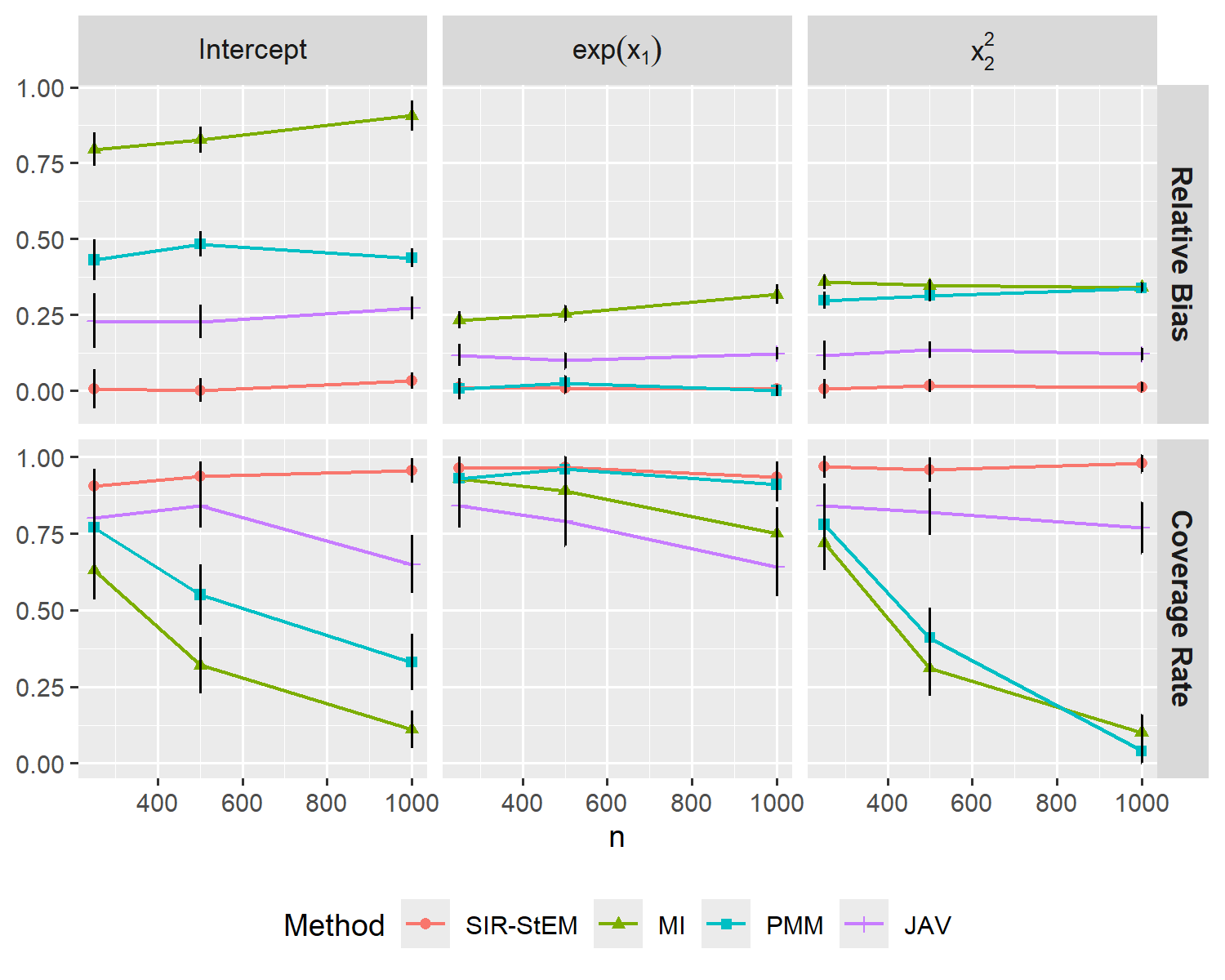} \\
\caption{Average relative bias and coverage rates by $n$ and method.
Error bars indicate $\pm 1$ standard error.}
\label{fig:study1_main}
\end{figure}

To begin we focus on the most challenging conditions of $\varphi_{\mathrm{MIS}} = 0.3$ and $R^2 = 0.50$.
Plots of mean relative bias and coverage rates for this condition are displayed in Figure~\ref{fig:study1_main}.
Across all parameters, SIR-StEM exhibits negligible relative bias that remains stable as $n$ increases.
In contrast, MI displays substantial relative bias, exceeding 0.75 for the intercept and about 0.25 for the other coefficients.
PMM shows near-zero bias for $\exp(X_1)$ but about 0.45 for the intercept and above 0.30 for $X_2^2$ across all $n$.
JAV is generally less biased than MI and PMM, but still exhibits at least 0.20 relative bias for the intercept, about 0.10 for $\exp(X_1)$, and about 0.10 for $X_2^2$ for all sample sizes.
For coverage, SIR-StEM maintains rates close to the nominal 95\% level for all parameters and sample sizes, with the exception of minor undercoverage in the intercept when $n = 250$.
In contrast, MI, PMM, and JAV suffer from undercoverage that worsens as $n$ increases, particularly for the intercept and $X_2^2$.
An exception is PMM on the $\exp(X_1)$ parameter, which retains near-nominal coverage alongside its low bias.

The other conditions that include $\varphi_{\mathrm{MIS}} \in \{0.1, 0.2\}$ with $R^2 \in \{0.25, 0.5\}$ displayed similar patterns of results and have been relegated to Appendix~\ref{app:supp_results} for brevity.
Briefly, the notable trends are that reducing $\varphi_{\mathrm{MIS}}$ uniformly improves bias and coverage slightly for MI, PMM, and JAV, but they generally remain systematically inaccurate.
SIR-StEM remains stable across all missingness levels, with coverage largely insensitive to $\varphi_{\mathrm{MIS}}$.
Lowering $R^2$ attenuated bias and undercoverage for MI, PMM, and JAV, having a notably larger effect on JAV than MI and PMM.
SIR-StEM continued to exhibit both low bias and near-nominal coverage in these settings.

Across all combinations of signal strength, missingness, 
and sample size, SIR-StEM is the only method that 
consistently achieves low bias and near-nominal coverage.
The other methods remain biased to varying degrees, with MI exhibiting the most bias and JAV exhibiting the least.
Moreover, these methods displayed undercoverage that worsens with sample size (aside from PMM on $\exp(X_1)$).
These patterns are generally preserved at lower missingness and $R^2$, though reducing either of these factors improves the performance of these methods.

\subsection{Nonparametric Transformation of Behavioral Predictor}

In this study, we compare missing data methods in a nonlinear regression setting where the conditional mean is estimated through spline basis functions.
We use a predictor generated by nonparametric bootstrap from a real dataset to emulate data characteristics found in real psychological studies.
Conditional on this predictor, outcomes are generated from a fixed quadratic regression model with additive Gaussian noise, and missingness is introduced artificially.
This design allows us to study missingness in spline predictors that may be generated from realistic psychological data.

We analyzed measures of psychopathology from the Adolescent Brain Cognitive Development (ABCD) Study (\url{https://abcdstudy.org}).
The ABCD Study is a large, multi-site study, whose data are publicly available \citep{Volkow2018}, which was approved by the institutional review boards of the participating sites \citep{Clark2018}.
To avoid potential clustering effects by site and to reduce the sample size to a more realistic scale, one site was selected randomly to provide the basis of our data ($n = 631$).

\begin{figure}[!th]
\centering
\includegraphics[width=0.9\textwidth, keepaspectratio]{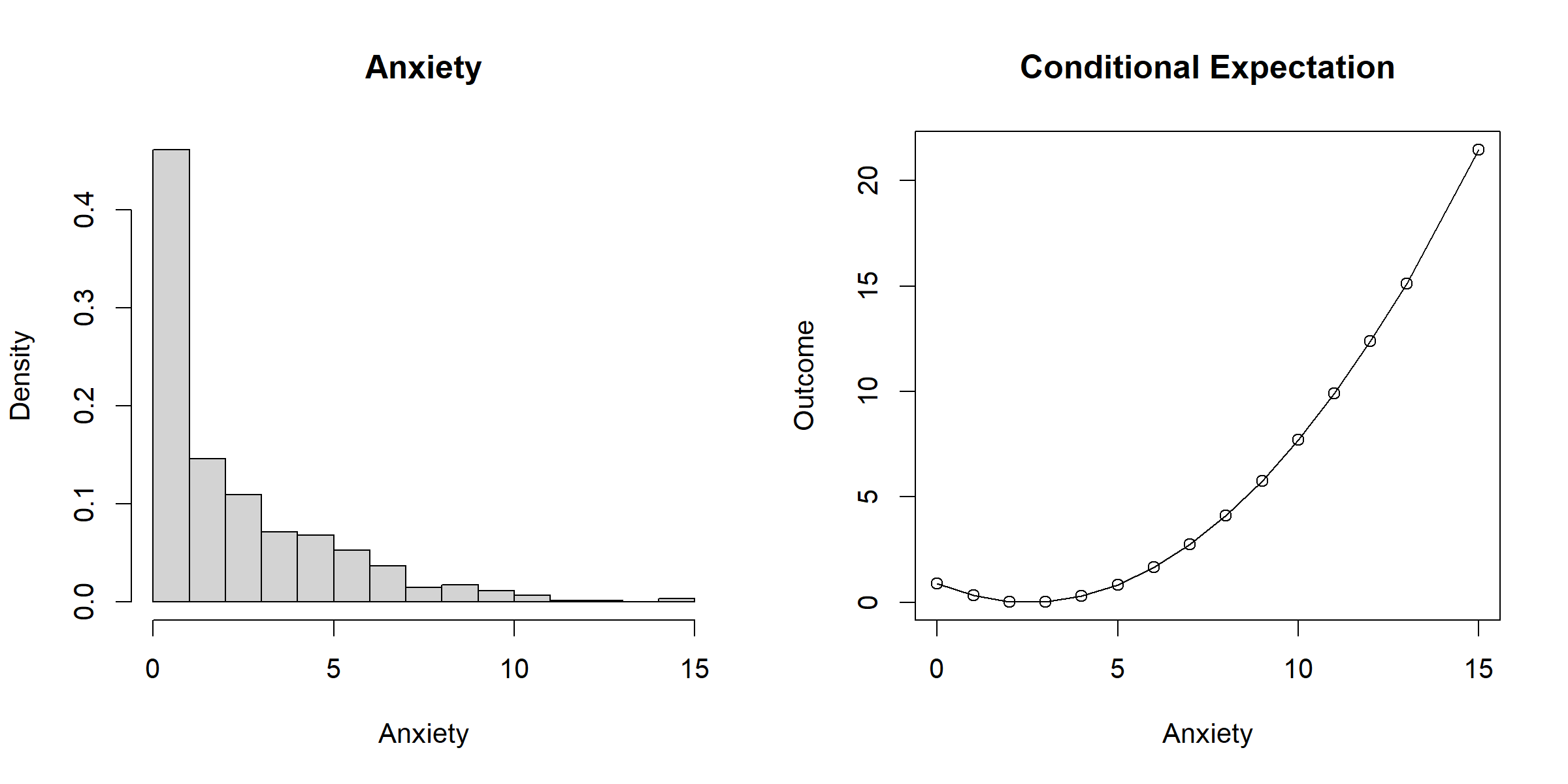} \\
\caption{Histograms of anxiety measure (left). Conditional expectation of the outcome (right).}
\label{fig:study2_desc}
\end{figure}

Measures were taken using summary scores of the Child Behavior Checklist \citep[CBCL;][]{Achenbach2001}.
These scores are fairly skewed and discrete, residing on the integers 0 to 15.
The empirical distribution of anxiety scores served as the population of the predictors, where the observed proportion was taken to be the probability of each score in the data-generating process.
We note that the SIR-StEM algorithm uses a Gaussian assumption for $X$, thus this design serves as a robustness check under misspecification of the predictor distribution.
We considered a quadratic model of an outcome as a function of anxiety as follows
\begin{equation}
f(X) = \dfrac{(X - \E[X])^2}{\sqrt{\Var(X)}},
\end{equation}
with additive errors
\begin{equation}
Y = f(X) + \epsilon,
\end{equation}
where $\epsilon \sim \mathcal{N}(0, \sigma^2)$, and $\sigma^2$ was chosen such that $R^2 = 0.25$ as in the previous simulation.
We display histograms of these scores and a plot of the conditional expectation function in Figure~\ref{fig:study2_desc}.
For each bootstrapped dataset, missingness was inserted using the same MAR generating procedure as the previous simulation study, fixing the proportion of missingness as $\varphi_{\mathrm{MIS}} = 0.3$.
We used natural cubic splines to estimate $f(x)$, and ran $r = 100$ simulation replications.

\begin{figure}[!th]
\centering
\includegraphics[width=0.9\textwidth, keepaspectratio]{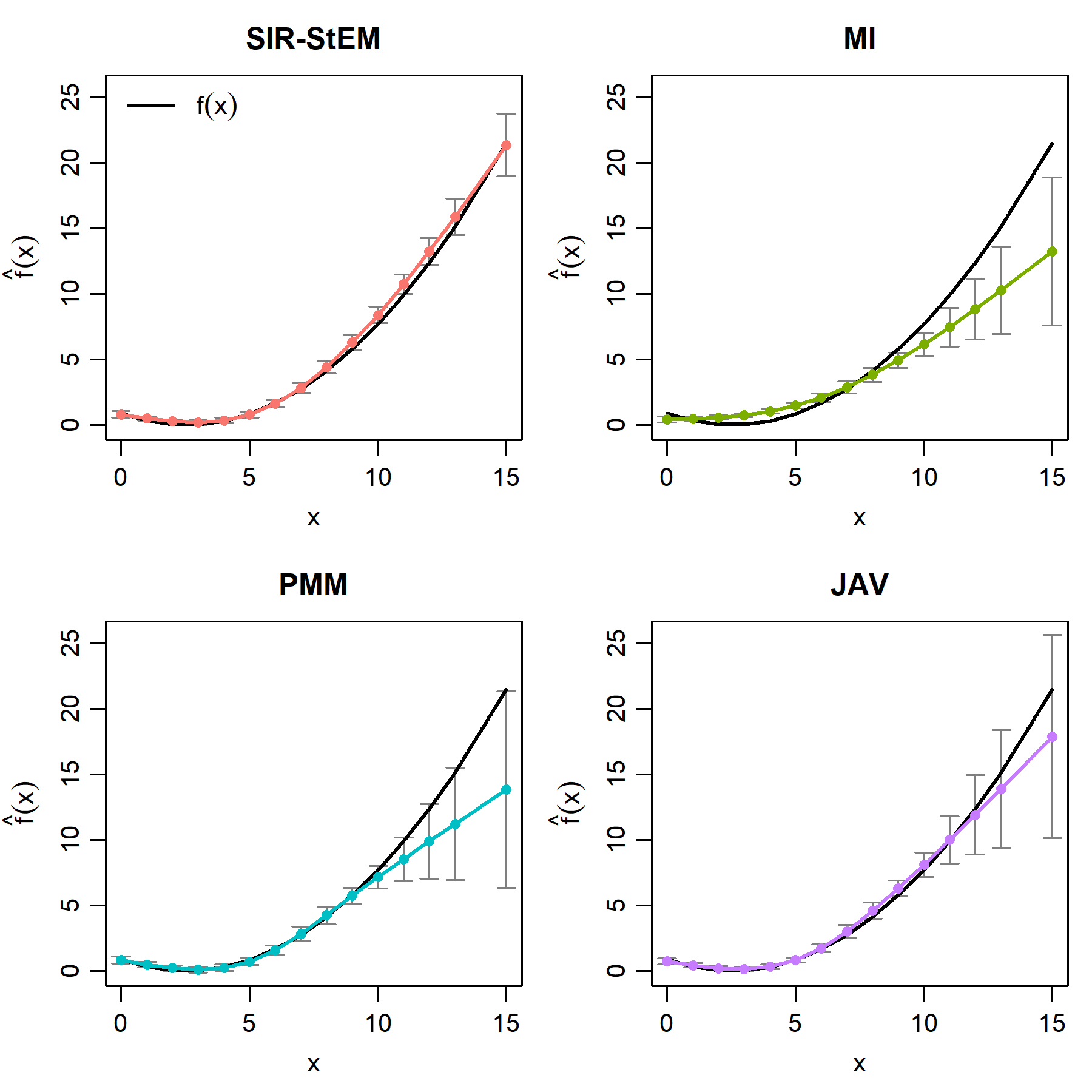} \\
\caption{Estimated $\widehat{f}(x)$ by method. The solid black line denotes the true $f(x)$ and the error bars indicate one standard deviation of $\widehat{f}(x)$.}
\label{fig:study2}
\end{figure}

We report Monte Carlo estimates of integrated mean squared error and integrated variance (IMSE and IVAR), along with an integrated 95\% coverage rate (ICOV):
\begin{equation}
\begin{aligned}
\text{IMSE} &\coloneqq \dfrac{1}{r} \sum_{i = 1}^r\, \sum_{x = 0}^{15} \left( \widehat{f}_i(x) - f(x) \right)^2 \Prob(x)\\
\text{IVAR} &\coloneqq \dfrac{1}{r} \sum_{i = 1}^r\, \sum_{x = 0}^{15} \left( \widehat{f}_i(x) - \bar{f}(x) \right)^2 \Prob(x)\\
95\%\,\text{ICOV} &\coloneqq \dfrac{1}{r} \sum_{i = 1}^r\, \sum_{x = 0}^{15} I\left(\lvert\widehat{f}_i(x) - f(x)\rvert \leq \Phi^{-1}(0.975) \sqrt{\widehat{\Var}(\hat{f}_i(x))}\right)\Prob(x)
\end{aligned}
\end{equation}
where $\bar{f}(x) = \sum_{i=1}^r \widehat{f}_i(x)/r$.
These estimates are displayed in Table~\ref{tab:study2_est}, and Figure~\ref{fig:study2} shows the Monte Carlo mean of $\widehat{f}(x)$ with its standard deviation across $x$ per method.

\begin{table}[!h]
\centering
\caption{Estimated IMSE, IVAR, and 95\% ICOV.}
\label{tab:study2_est}
\begin{tabular}{cSSS}
\toprule
Method & {$\text{IMSE}$} & {$\text{IVAR}$} & {$95\%\,\text{ICOV}$}\\
\midrule
SIR-StEM &  0.120 &  0.083 & 0.962\\
MI       &  0.790 &  0.202 & 0.647\\
PMM      &  0.569 &  0.327 & 0.946\\
JAV      &  0.390 &  0.323 & 0.985\\
\bottomrule
\end{tabular}
\end{table}

Figure~\ref{fig:study2} shows that SIR-StEM recovers $f(x)$ well over all $x$, with uniformly small variability, even though the Gaussian assumption for $X$ is misspecified.
MI and PMM follow $f(x)$ at low values of $x$, but increasingly underestimate the curve with larger variability as $x$ increases.
JAV recovers more of the upward trend than MI or PMM, but still also underestimates $f(x)$ with larger variability as $x$ increases.
These results are consistent with Table~\ref{tab:study2_est}, where SIR-StEM obtains the lowest IMSE and IVAR with near-nominal coverage, while MI shows the largest IMSE and severe undercoverage.
PMM and JAV show improvement over MI in IMSE and coverage, but remain well above SIR-StEM in both IMSE and IVAR.

\section{Discussion}

In this research, we studied a stochastic EM algorithm with sampling-importance resampling (SIR-StEM) for generalized linear regression models with missing data and nonlinear transformations of predictors.
The algorithm applies a direct sampling approach when the conditional distributions are tractable, and utilizes SIR sampling when they are not.
We also showed that the resulting estimators are asymptotically normal provided the number of proposal draws $n_q$ grows sufficiently quickly with the sample size, so that SIR-StEM inherits the limiting distribution of the StEM procedure with exact conditional draws.
Our two empirical simulations also showed that the SIR-StEM algorithm can perform well for some common models that are challenging for multiple imputation methods.

There are some limitations of the SIR-StEM algorithm to consider.
First, there may be difficulty in computational scaling with the number of substantive predictors $p$.
Since SIR may need to characterize a higher dimensional conditional distribution, larger amounts of proposal samples may be needed for effective resampling weights.
Second, the standard error procedure for SIR-StEM is also somewhat computationally demanding.
It entails a parametric bootstrap to estimate the observed-data information matrix, then a batch-mean procedure on a Markov chain.
This is in contrast to the computational simplicity of Rubin's rules that multiple imputation procedures use.
Third, the Gaussian model for $X$ is a working assumption.
Our second simulation showed that SIR-StEM can still perform well when this assumption is violated by skewed and discrete predictors, though categorical predictors are not explicitly handled by the algorithm.

For future directions, improved or adaptive proposal distributions could reduce the required $n_q$ and improve computational efficiency in higher dimensions.
Extensions to the regression model may also be considered, which include categorical predictors, mixed-effects models, and regularized regression models.
These extensions may require adaptations to the sampling scheme and maximization steps.
Finally, an accessible software implementation would help make the method more broadly usable.

\bibliographystyle{asa}
\bibliography{references} 

\newpage
\appendix
\appendixpage
\section{Asymptotic Normality of the SIR-StEM Algorithm} \label{app:stemsir}

Let $\widetilde{U}^{(t)}_n$ be $n$ complete-data samples from $\Prob_{\theta^{(t)}}(U)$.
Also let $\widetilde{U}^{(t)}_{n_q, n}$ be $n$ complete-data samples from $\Prob_{\theta^{(t)}, n_q}(U)$ which is the SIR approximation to $\Prob_{\theta^{(t)}}(U)$ using $n_q$ proposals.
First, by convergence in distribution of the SIR draw in $n_q$ \citep{Smith1992}, we have $\widetilde{U}^{(t)}_{n_q, n} \xrightarrow{d} \widetilde{U}^{(t)}_n$ as $n_q \rightarrow \infty$, for any fixed $n$.

Now let $\theta \in \mathbb{R}^d$ and let the $M$-step be a continuous map $M:\mathbb{R}^{n \times (p + 1)} \times \mathbb{R}^d \rightarrow \mathbb{R}^d$ such that $\theta^{(t + 1)} = M(\widetilde{U}^{(t)}, \theta^{(t)})$.
We may subsequently denote $\theta^{(t + 1)}_n = M(\widetilde{U}^{(t)}_n, \theta^{(t)}_n)$ and $\theta^{(t + 1)}_{n_q, n} = M(\widetilde{U}^{(t)}_{n_q, n}, \theta^{(t)}_{n_q, n})$.
We will prove that $\theta^{(t)}_{n_q, n} \xrightarrow{d} \theta^{(t)}_n$ as $n_q \rightarrow \infty$ for a fixed $n$ and all $t$ by induction.

First we prove the base case.
Let $\theta^{(0)}$ be a fixed constant.
Then we have
\begin{equation}
\begin{aligned}
M(\widetilde{U}^{(0)}_{n_q, n}, \theta^{(0)}) &\xrightarrow{d} M(\widetilde{U}^{(0)}_n, \theta^{(0)})\\
\theta^{(1)}_{n_q, n} &\xrightarrow{d} \theta^{(1)}_n
\end{aligned}
\end{equation}
by the continuous mapping theorem.
Next we prove the induction step.
Assume that $\theta^{(t)}_{n_q, n} \xrightarrow{d} \theta^{(t)}_n$ as the induction hypothesis, then we have
\begin{equation}
\begin{aligned}
M(\widetilde{U}^{(t)}_{n_q, n}, \theta^{(t)}_{n_q, n}) &\xrightarrow{d} M(\widetilde{U}^{(t)}_n, \theta^{(t)}_{n})\\
\theta^{(t+1)}_{n_q, n} &\xrightarrow{d} \theta^{(t+1)}_n
\end{aligned}
\end{equation}
by another application of the continuous mapping theorem, together with the assumed continuous convergence of $\Prob_{\theta, n_q}(U)$ to $\Prob_{\theta}(U)$.
Hence we have $\theta^{(t)}_{n_q, n} \xrightarrow{d} \theta^{(t)}_n$ as $n_q \rightarrow \infty$ for any fixed $n$ and $t$.

Now, denote the asymptotically scaled StEM and SIR-StEM estimates as
\begin{equation}
\begin{aligned}
Z^{\mathrm{SIR}}_{n_q, n} &\coloneqq \sqrt{n}(\thetasir - \theta_0)\\
Z^{\mathrm{StEM}}_n &\coloneqq \sqrt{n}(\thetastem - \theta_0),
\end{aligned}
\end{equation}
which are continuous functions of the iterates $\theta^{(t)}_{n_q, n}$ and $\theta^{(t)}_n$, respectively. 
By the continuous mapping theorem, for each fixed $n$ we then have $Z^{\mathrm{SIR}}_{n_q, n} \xrightarrow{d} Z^{\mathrm{StEM}}_n$ as $n_q \rightarrow \infty$.

We now establish it is possible to maintain this convergence as $n \rightarrow \infty$.
This can be shown with any metric $d(A, B)$, which refers to a distance on the distributions of the random variables $A$ and $B$, and where $d(\cdot, \cdot)$ additionally metrizes convergence in distribution (e.g., L\'evy-Prokhorov metric).
Then $Z^{\mathrm{SIR}}_{n_q, n} \xrightarrow{d} Z^{\mathrm{StEM}}_n$ as $n_q \rightarrow \infty$ implies
\begin{equation} \label{eq:lp_pointwise}
d\left(Z^{\mathrm{SIR}}_{n_q, n}, Z^{\mathrm{StEM}}_n\right) \rightarrow 0 \quad \text{as } n_q \rightarrow \infty.
\end{equation}

Let us consider an error tolerance for this limit as $1/n$.
Then there exists a threshold $N_0(n)$ such that the distance is below $1/n$.
We then create a nondecreasing function in $n$ by setting $N(n) = \max\{N_0(1), \ldots, N_0(n), n\}$, letting $n_q(n)$ be any sequence such that $n_q(n) \geq N(n)$ for all $n$.
Thus there exists an $n_q(n) \geq N(n)$ such that $d\left(Z^{\mathrm{SIR}}_{n_q(n), n}, Z^{\mathrm{StEM}}_n\right) < 1/n$.
This implies
\begin{equation} \label{eq:d_in_n}
d\left(Z^{\mathrm{SIR}}_{n_q(n), n}, Z^{\mathrm{StEM}}_n\right) \rightarrow 0 \quad \text{as } n \rightarrow \infty,
\end{equation}
under an appropriate construction of $n_q(n)$.

Now, under the regularity conditions of \citet{Nielsen2000}, we have $Z^{\mathrm{StEM}}_n \xrightarrow{d} \mathcal{N}(0, \Gamma)$, where $\Gamma$ is the asymptotic covariance of $\thetastem$.
Then using the triangle inequality we have
\begin{equation} \label{eq:triangle}
d\left(Z^{\mathrm{SIR}}_{n_q(n), n}, \mathcal{N}(0, \Gamma)\right) \leq d\left(Z^{\mathrm{SIR}}_{n_q(n), n}, Z^{\mathrm{StEM}}_{n}\right) + d\left(Z^{\mathrm{StEM}}_{n}, \mathcal{N}(0, \Gamma)\right)
\end{equation}
with both terms on the right-hand side converging to zero as $n \rightarrow \infty$.
This ultimately yields the result
\begin{equation}
\sqrt{n}(\widehat{\theta}^{\mathrm{SIR}}_{n_q(n), n} - \theta_0) \xrightarrow{d} \mathcal{N}(0, \Gamma)
\end{equation}
as $n \rightarrow \infty$.
If we use the average of a length $m$ Markov chain as the estimator $\thetasir$, then $\Gamma = \mathcal{I}(\theta_0)^{-1} + \Gamma_{m}$, as in Proposition~5 of \citet{Nielsen2000}.

\section{Additional Results} \label{app:supp_results}

\begin{figure}[!th]
\centering
\includegraphics[width=\textwidth, keepaspectratio]{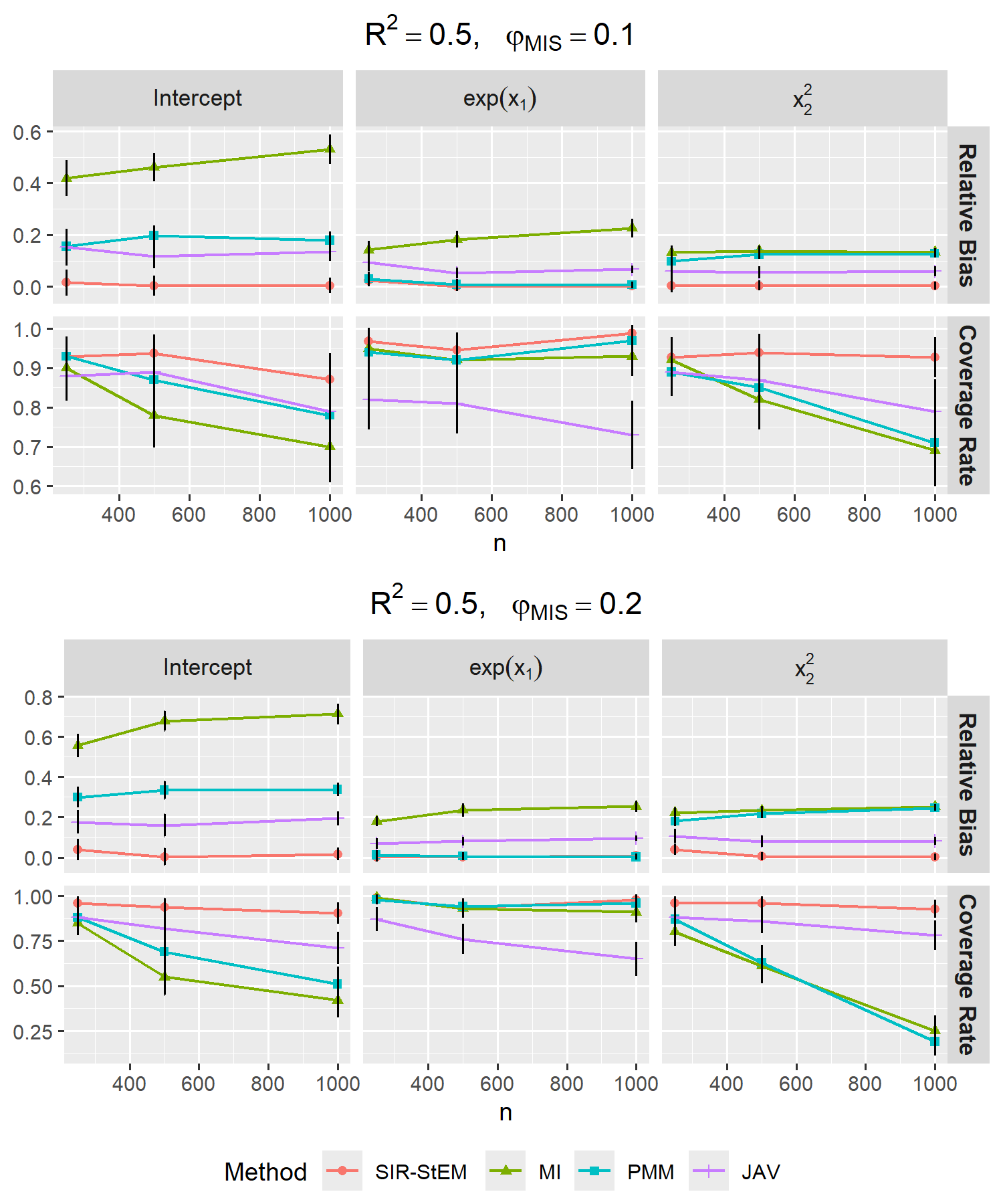} \\
\caption{Average relative bias and coverage rates by $n$, $\varphi_{\mathrm{MIS}}$ and method.
Error bars indicate $\pm 1$ standard error.}
\label{fig:study1_app1}
\end{figure}

\begin{figure}[!th]
\centering
\includegraphics[width=\textwidth, keepaspectratio]{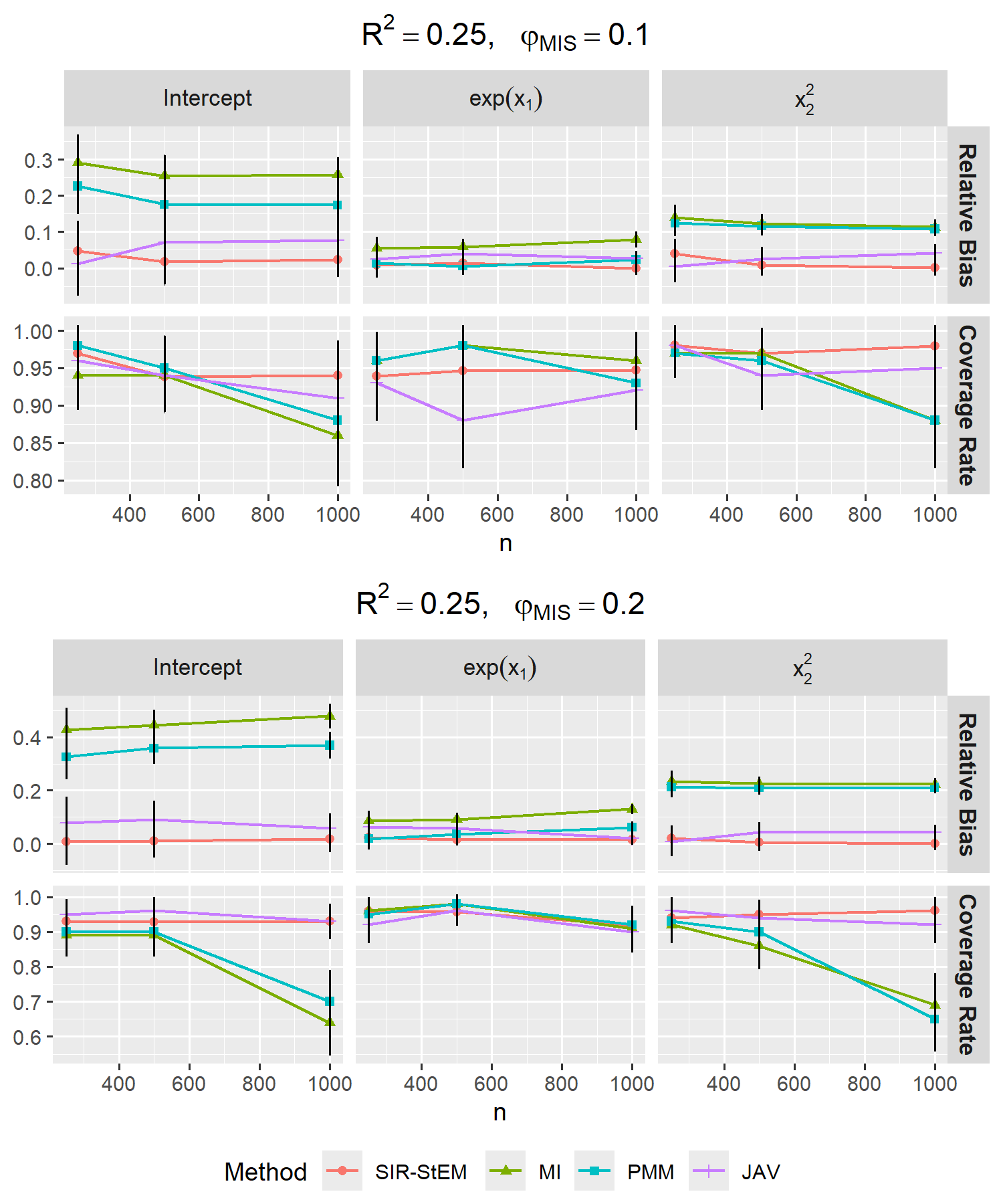} \\
\caption{Average relative bias and coverage rates by $n$, $\varphi_{\mathrm{MIS}}$ and method.
Error bars indicate $\pm 1$ standard error.}
\label{fig:study1_app2}
\end{figure}

\begin{figure}[!th]
\centering
\includegraphics[width=\textwidth, keepaspectratio]{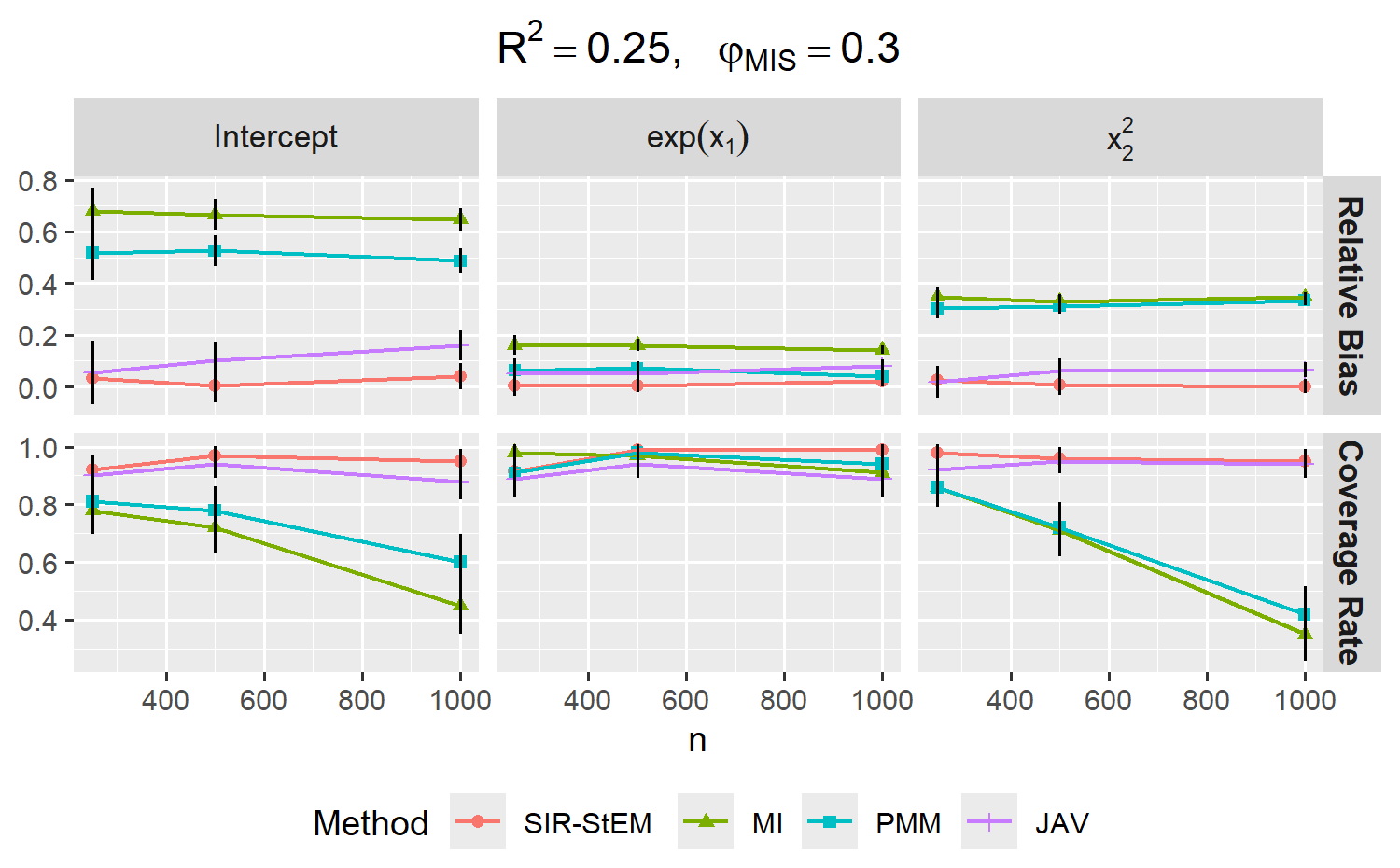} \\
\caption{Average relative bias and coverage rates by $n$ and method.
Error bars indicate $\pm 1$ standard error.}
\label{fig:study1_app3}
\end{figure}

We display the additional results in Figures~\ref{fig:study1_app1} through~\ref{fig:study1_app3}.
These conditions are the remaining combinations that include $\varphi_{\mathrm{MIS}} \in \{0.1, 0.2\}$ and $R^2 \in \{0.25, 0.5\}$.

\end{document}

%% file: style.tex
\setlist{noitemsep, topsep=0pt}
\bibpunct{(}{)}{;}{a}{}{,} 
\usepackage[english]{babel}
\usepackage[utf8]{inputenc}
\usepackage{fancyhdr}
 


%% file: references.bib
@article{Rubin1976,
author = {Rubin, Donald B.},
journal = {Biometrika},
number = {3},
pages = {581--592},
title = {{Inference and missing data}},
volume = {63},
year = {1976}
}

@book{Schafer1997,
address = {Boca Raton, FL},
author = {Schafer, J. L.},
publisher = {Chapman {\&} Hall},
title = {{Analysis of Incomplete Multivariate Data}},
year = {1997}
}

@article{Zhang2017,
author = {Zhang, Qian and Wang, Lijuan},
journal = {Psychological Methods},
number = {4},
pages = {649--666},
title = {{Moderation analysis with missing data in the predictors}},
volume = {22},
year = {2017}
}

@article{Ludtke2019,
author = {L{\"{u}}dtke, Oliver and Robitzsch, Alexander and West, Stephen G.},
journal = {Multivariate Behavioral Research},
title = {{Analysis of interactions and nonlinear effects with missing data: A factored regression modeling approach using maximum likelihood estimation}},
volume = {55},
number = {3},
pages = {361--381},
year = {2020}
}

@article{Kim2015,
author = {Kim, Soeun and Sugar, Catherin A. and Belin, Thomas R.},
journal = {Statistics in Medicine},
number = {11},
pages = {1876--1888},
title = {{Evaluating model-based imputation methods for missing covariates in regression models with interactions}},
volume = {34},
year = {2015}
}

@article{Seaman2012,
author = {Seaman, Shaun R. and Bartlett, Jonathan W. and White, Ian R.},
journal = {BMC Medical Research Methodology},
number = {46},
pages = {1--13},
title = {{Multiple imputation of missing covariates with non-linear effects and interactions: An evaluation of statistical methods}},
volume = {12},
year = {2012}
}

@article{Bartlett2015,
author = {Bartlett, Jonathan W. and Seaman, Shaun R. and White, Ian R. and Carpenter, James R.},
journal = {Statistical Methods in Medical Research},
number = {4},
pages = {462--487},
title = {{Multiple imputation of covariates by fully conditional specification: Accommodating the substantive model}},
volume = {24},
year = {2015}
}

@article{VonHippel2009,
author = {von Hippel, Paul T.},
journal = {Sociological Methodology},
number = {1},
pages = {265--291},
title = {{How to impute interactions, squares, and other transformed variables}},
volume = {39},
year = {2009}
}

@article{Cowles1996,
author = {Cowles, Mary Kathryn and Carlin, Bradley P.},
journal = {Journal of the American Statistical Association},
number = {434},
pages = {883--904},
title = {{Markov chain Monte Carlo convergence diagnostics: A comparative review}},
volume = {91},
year = {1996}
}

@article{Hinrichs2014,
author = {Hinrichs, A. and Novak, E. and Ullrich, M. and Wo{\'{z}}niakowski, H.},
journal = {Mathematics of Computation},
pages = {2853--2863},
title = {{The curse of dimensionality for numerical integration of smooth functions}},
volume = {83},
year = {2014}
}

@article{Simonovits2003,
author = {Simonovits, Mikl{\'{o}}s},
journal = {Mathematical Programming},
pages = {337--374},
title = {{How to compute the volume in high dimension?}},
volume = {97},
year = {2003}
}

@article{Dempster1977,
author = {Dempster, A. P. and Laird, Nan M. and Rubin, Donald B.},
journal = {Journal of the Royal Statistical Society: Series B},
number = {1},
pages = {1--38},
title = {{Maximum likelihood from incomplete data via the EM algorithm}},
volume = {39},
year = {1977}
}

@article{Wu1983,
author = {Wu, C. F. Jeff},
journal = {The Annals of Statistics},
number = {1},
pages = {95--103},
title = {{On the convergence properties of the EM algorithm}},
volume = {11},
year = {1983}
}

@article{Meng1991,
author = {Meng, Xiao-Li and Rubin, Donald B.},
journal = {Journal of the American Statistical Association},
number = {416},
pages = {899--909},
title = {{Using EM to obtain asymptotic variance-covariance matrices: The SEM Algorithm}},
volume = {86},
year = {1991}
}

@article{Jamshidian2000,
author = {Jamshidian, Mortaza and Jennrich, Robert I.},
journal = {Journal of the Royal Statistical Society: Series B (Statistical Methodology)},
number = {2},
pages = {257--270},
title = {{Standard errors for EM estimation}},
volume = {62},
year = {2000}
}

@article{Volkow2018,
author = {Volkow, Nora D. and Koob, George F. and Croyle, Robert T. and Bianchi, Diana W. and Gordon, Joshua A. and Koroshetz, Walter J. and P{\'{e}}rez-stable, Eliseo J. and Riley, William T. and Bloch, Michele H. and Conway, Kevin and Deeds, Bethany G. and Dowling, Gayathri J. and Grant, Steven and Howlett, Katia D. and Matochik, John A. and Morgan, Glen D. and Murray, Margaret M. and Noronha, Antonio and Spong, Catherine Y. and Wargo, Eric M. and Warren, Kenneth R. and Weiss, Susan R. B.},
journal = {Developmental Cognitive Neuroscience},
pages = {4--7},
title = {{The conception of the ABCD study : From substance use to a broad NIH collaboration}},
volume = {32},
year = {2018}
}

@article{Clark2018,
author = {Clark, Duncan B. and Fisher, Celia B. and Bookheimer, Susan and Brown, Sandra A. and Evans, John H. and Hopfer, Christian and Hudziak, James and Montoya, Ivan and Murray, Margaret and Pfefferbaum, Adolf and Yurgelun-Todd, Deborah},
journal = {Developmental Cognitive Neuroscience},
pages = {143--154},
title = {{Biomedical ethics and clinical oversight in multisite observational neuroimaging studies with children and adolescents: The ABCD experience}},
volume = {32},
year = {2018}
}

@book{Achenbach2001,
address = {Burlington, VT},
author = {Achenbach, T. M. and Rescorla, L. A.},
publisher = {University of Vermont},
title = {{Manual for the ASEBA school-age forms {\&} profiles: An integrated system of multi-informant assessment}},
year = {2001}
}

@article{Ren2022, title={{Nonlinear effect of social interaction quantity on psychological well-being: Diminishing returns or inverted U?}}, volume={122}, number={6}, journal={Journal of Personality and Social Psychology}, publisher={American Psychological Association}, author={Ren, Dongning and Stavrova, Olga and Loh, Wen Wei}, year={2022}, pages={1056} }

@article{Verhaeghen1997, title={Meta-analyses of age–cognition relations in adulthood: Estimates of linear and nonlinear age effects and structural models.}, volume={122}, number={3}, journal={Psychological Bulletin}, publisher={American Psychological Association}, author={Verhaeghen, Paul and Salthouse, Timothy A}, year={1997}, pages={231} }

@article{Watters1997, title={{Caffeine and cognitive performance: The nonlinear Yerkes–Dodson law}}, volume={12}, number={3}, journal={Human Psychopharmacology: Clinical and Experimental}, publisher={Wiley Online Library}, author={Watters, Paul Andrew and Martin, Frances and Schreter, Zoltan}, year={1997}, pages={249–257} }

@article{Yerkes1908, title={The relation of strength of stimulus to rapidity of habit‐formation}, volume={18}, rights={http://onlinelibrary.wiley.com/termsAndConditions#vor}, ISSN={0092-7015, 1550-7149}, DOI={10.1002/cne.920180503}, number={5}, journal={Journal of Comparative Neurology and Psychology}, author={Yerkes, Robert M. and Dodson, John D.}, year={1908}, month=nov, pages={459–482}, language={en} }

@article{Kim2026, title={A Hybrid {EM} Algorithm for Linear Two-Way Interactions With Missing Data}, volume={51}, ISSN={1076-9986, 1935-1054}, DOI={10.3102/10769986241304015}, number={1}, journal={Journal of Educational and Behavioral Statistics}, author={Kim, Dale S.}, year={2026}, month=feb, pages={38–59}, language={en} }

@article{Nielsen2000, title={The stochastic {EM} algorithm: Estimation and asymptotic results}, volume={6}, ISSN={13507265}, DOI={10.2307/3318671}, number={3}, journal={Bernoulli}, author={Nielsen, Søren Feodor}, year={2000}, month=jun, pages={457} }

@article{Celeux1985, title={The {SEM} algorithm: A probabalistic teacher algorithm derived from the {EM} algorithm for the mixture problem}, volume={2}, journal={Computational Statistics Quarterly}, author={Celeux, G. and Diebolt, J.}, year={1985}, pages={73–82} }

@incollection{Diebolt1995,
  author = {Diebolt, Jean and Ip, Eddie H. S.},
  title = {Stochastic {EM}: Method and application},
  booktitle = {Markov Chain Monte Carlo in Practice},
  editor = {Gilks, W. R. and Richardson, S. and Spiegelhalter, D. J.},
  publisher = {CRC Press},
  year = {1995},
  pages = {259--273}
}

@book{Gelman2014, address={Boca Raton}, edition={Third edition}, series={Chapman {\&} Hall/CRC Texts in Statistical Science}, title={Bayesian Data Analysis}, ISBN={978-1-4398-4095-5}, callNumber={QA279.5 .G45 2014}, publisher={CRC Press}, author={Gelman, Andrew and Carlin, John B. and Stern, Hal S. and Dunson, David B. and Vehtari, Aki and Rubin, Donald B.}, year={2014}, collection={Chapman & Hall/CRC Texts in Statistical Science}}

@book{Liu2008, address={New York, NY}, series={Springer Series in Statistics Ser}, title={{Monte Carlo Strategies in Scientific Computing}}, ISBN={978-0-387-76369-9}, publisher={Springer New York}, author={Liu, Jun S.}, year={2008}, collection={Springer Series in Statistics Ser}, language={eng}}

@article{Rubin1987, title={{A noniterative sampling/importance resampling alternative to the data augmentation algorithm for creating a few imputations when fractions of missing information are modest: The SIR algorithm}}, volume={82}, ISSN={0162-1459, 1537-274X}, DOI={10.1080/01621459.1987.10478461}, number={398}, journal={Journal of the American Statistical Association}, author={Rubin, Donald B.}, year={1987}, month=jun, pages={543–546}, language={en}}

@article{Smith1992, title={Bayesian statistics without tears: A sampling–resampling perspective}, volume={46}, ISSN={0003-1305, 1537-2731}, DOI={10.1080/00031305.1992.10475856}, number={2}, journal={The American Statistician}, author={Smith, A. F. M. and Gelfand, A. E.}, year={1992}, month=may, pages={84–88}, language={en} }

@book{Rubin1987b, address={New York}, series={Wiley series in probability and mathematical statistics Applied probability and statistics}, title={Multiple imputation for nonresponse in surveys}, ISBN={9780471087052}, DOI={10.1002/9780470316696}, publisher={Wiley}, author={Rubin, Donald B.}, year={1987}, collection={Wiley series in probability and mathematical statistics Applied probability and statistics}, language={eng} }

@article{Jones2006, title={{Fixed-width output analysis for Markov Chain Monte Carlo}}, volume={101}, ISSN={0162-1459, 1537-274X}, DOI={10.1198/016214506000000492}, number={476}, journal={Journal of the American Statistical Association}, author={Jones, Galin L and Haran, Murali and Caffo, Brian S and Neath, Ronald}, year={2006}, month=dec, pages={1537–1547}, language={en} }

@article{Louis1982, title={Finding the observed information matrix when using the {EM} algorithm}, volume={44}, ISSN={0035-9246, 2517-6161}, DOI={10.1111/j.2517-6161.1982.tb01203.x}, number={2}, journal={Journal of the Royal Statistical Society: Series B (Methodological)}, author={Louis, Thomas A.}, year={1982}, month=jan, pages={226–233}, language={en} }

@article{Geweke1989, title={{Bayesian Inference in Econometric Models Using Monte Carlo Integration}}, volume={57}, ISSN={00129682}, DOI={10.2307/1913710}, number={6}, journal={Econometrica}, author={Geweke, John}, year={1989}, month=nov, pages={1317} }
